\documentclass[aps,twocolumn,showpacs]{revtex4}
\usepackage{amsmath,times}
\usepackage{graphicx,subfigure}
\usepackage{hyperref}

\begin{document}

\title{Classical field techniques for condensates in one-dimensional rings at finite temperatures}
\author{A.~Nunnenkamp, J.~N.~Milstein and K.~Burnett}
\affiliation{Clarendon Laboratory, Department of Physics, University
of Oxford, Parks Road, Oxford OX1 3PU, United Kingdom}

\begin{abstract}
For a condensate in a one-dimensional ring geometry, we compare the thermodynamic properties of three conceptually different classical field techniques: stochastic dynamics, microcanonical molecular dynamics, and the ``classical field method.''  Starting from non-equilibrium initial conditions, all three methods approach steady states whose distribution and correlation functions are in excellent agreement with an exact evaluation of the partition function in the high-temperature limit.  Our study helps to establish these various classical field techniques as powerful non-perturbative tools for systems at finite temperatures.
\end{abstract}


\maketitle

\section{Introduction}

Future experiments on ultracold gases will require theoretical input from numerical simulations which are reliable in the case of strong interactions beyond the realm of perturbative expansions.  Powerful non-perturbative techniques have been developed ranging from the partial summation of Feynman diagrams in the context of condensed matter and nuclear theory (e.g.~the many-body T-matrix approach \cite{Mahan90}) to more recent renormalization group ideas inspired from quantum information theory (e.g.~the density-matrix renormalization group (DMRG) \cite{Schollwock05}). These methods, however, are often intractable in systems at finite temperatures or beyond one spatial dimension.

Many of the complications at finite temperatures arise from a need to account for higher-order classical fluctuations, for instance, near a second-order phase transition.  Here, a rather broad category of methods, which we will refer to as ``classical field'' (CF) methods, have proved to be a useful alternative.  Initially conceived of as a convenient way to sample the thermodynamic partition function, similar to Monte-Carlo techniques \cite{Negele90}, a variant of these CF methods, known as the ``classical field method'' \footnote{The unfortunate, yet consistent within the literature, naming of this particular classical field method as ``the classical field method,'' can lead to some confusion.  To stem this confusion, we will retain the quotes when referencing this particular method and refer to the broader group of methods as CF methods.}, has recently been developed and employed to study both the thermodynamics and real-time dynamics of cold atomic systems.  Some recent examples are studies of vortex-antivortex binding in the Kosterlitz-Thouless transition \cite{Simula06} or the shift of the Bose-Einstein condensation (BEC) transition temperature in an interacting Bose gas \cite{Davis06}. Despite its relative success, a rigorous, formal understanding of the level of fluctuations included by the ``classical field method,'' and its relation to more traditional CF methods, is still lacking.

In this paper, we establish that, in the case of an interacting Bose gas in a one-dimensional ring geometry, the ``classical field method'' approaches the same equilibrium as the more traditional stochastic and molecular dynamics CF methods, and that this equilibrium agrees with an exact evaluation of the partition function in the high-temperature limit. First, we discuss the stochastic dynamics approach which, in the flavor of the time-dependent Ginzburg-Landau equation, relaxes the field configuration to the minimum of the free energy while accounting for thermal fluctuations about this minimum.  Second, we evaluate the microcanonical dynamics method which resembles a molecular dynamics
approach based on Hamilton's equations of motion.  Third, we look at the ``classical field method'' which is based on the non-linear Schr\"odinger equation (NLSE).

In the case of the one-dimensional Bose gas on a ring, that we consider here, there is a simple way to evalutate the partition function at finite temperatures known as the transfer integral method. This solution serves us as a reference to which we can compare the properties of the steady states which the simulations reach. We find that the thermodynamic properties, such as distribution and correlation functions, of all three simulations, agree very well with the exact solution given by the transfer integral.

\section{The transfer integral solution}

We begin by evaluating the high-temperature limit of the partition function for an interacting Bose gas on a one-dimensional ring, exactly, by means of the transfer integral method \cite{Scalapino72}. This will serve as a reference by which we may judge the various CF methods discussed in this paper.

All information about the equilibrium properties of a many-body system may be extracted from its partition function $Z = \int D(\phi,\phi^*) e^{-S}$, where the action $S[\phi,\phi^*]$ reads
\begin{equation}
S[\phi,\phi^*] = \int_0^{\beta} d\tau  \int dx \, \left( \phi^*\partial_{\tau}
\phi + H[\phi,\phi^*] - \mu N[\phi,\phi^*] \right),
\end{equation}
$\mu$ is the chemical potential, and $\beta = 1/T$ denotes the reciprocal temperature (we set $k_B = 1$).

In the high temperature limit, $\beta \rightarrow 0$, the integration over imaginary time becomes quenched, and paths not on the $\tau = 0$ axis can be neglected, thus one arrives at what we will refer to as the ``classical partition function"
\begin{equation}
\label{classpf}
Z = \int D(\phi,\phi^*) \, e^{ - \beta F[\phi,\phi^*] },
\end{equation}
where we have introduced $F = \int dx \, (H - \mu N)$ the free energy. Although simpler than the full quantum-mechanical partition function, the classical partition function of Eq.~(\ref{classpf}) is, in general, a very difficult quantity to calculate. This is especially true near a second-order phase transition where all orders of classical fluctuations may contribute.

A Bose gas with contact interactions can be described by
\begin{equation}
F = \int dx \, \left[ \phi^* ( - \hbar^2\nabla^2/(2m) - \mu) \phi + g
|\phi|^4/2 \right],
\label{Bose_gas}
\end{equation}
where $g$ is the interaction constant and $m$ is the atomic mass. Its free energy is of the general form
\begin{equation}
F = \int dx \, \left( a |\phi|^2 + b |\phi|^4 + c |\nabla \phi|^2 \right).
\end{equation}
Defining a length scale $\xi_0 = \sqrt{c/a} = \hbar/\sqrt{2m\mu}$ which induces an energy scale $\epsilon_0 = \hbar^2/(2m\xi_0^2)$ we can rewrite the free energy in dimensionless quantities
\begin{equation}
\beta F = \tilde{\beta} \int d\tilde{x} \, \left( \tilde{a}
|\tilde{\phi}|^2 + \tilde{b} |\tilde{\phi}|^4 + |\tilde{\nabla}
\tilde{\phi}|^2 \right).
\label{free_energy}
\end{equation}
In the following we will leave off the tilde and all symbols refer to dimensionless quantities.

In one spatial dimension, the transfer integral method relates the classical partition function of Eq.~(\ref{classpf}), for a free energy of the form of Eq.~(\ref{free_energy}), to the Schr\"odinger equation of a single particle in an anharmonic oscillator potential
\begin{equation}
\label{transferintegral}
\left( -\frac{1}{4\beta^2} \nabla_u^2 + a |u|^2 + b |u|^4 \right)
\varphi_n(u) = E_n \varphi_n(u).
\end{equation}
In the appendix we show that thermodynamic properties, such as the distribution function of field amplitudes $P(|\phi| = u)$ and the two-point correlation function $\langle\phi(\bullet)^* \phi(\bullet+r)\rangle$, can readily be calculated from the eigenfunctions $\varphi_n(u)$ and eigenvalues $E_n$ of the single-particle Schr\"odigner equation, Eq.~(\ref{transferintegral}), by means of
\begin{equation}
P(|\phi| = u) = 2 u |\varphi_0(u)|^2,
\label{distribution}
\end{equation}
and
\begin{equation}
\begin{split}
\label{corr_function}
& \langle\phi(\bullet)^* \phi(\bullet+r)\rangle \\ = &
\sum_{n=1}^\infty \left| \int du \, \varphi_n^*(u) \, u \,
\varphi_0(u) \right|^2 e^{-\beta (|r|/\xi_0) (E_n-E_0) }.
\end{split}
\end{equation}
The Schr\"odinger equation of Eq.~(\ref{transferintegral}) can easily be solved numerically by exact diagonalization.  This approach provides a powerful way to evaluate the high-temperature limit of the partition function for an interacting system exactly which we can compare to our numerical simulations.

\section{Stochastic Dynamics}

An alternative method to evaluate the partition function of Eq.~(\ref{classpf}) is to allow the fields to evolve in such a way that they relax to the correct equilibrium distribution and to dynamically sample the relevant observables. This method of determining the equilibrium quantities was originally proposed by Parisi and Wu \cite{Parisi81}, while a related analysis has been used to model the behavior of a Fermi gas close to the BCS transition point \cite{Houbiers99}.

The fictional ``dynamics'' are determined by a diffusive term which relaxes the field configuration to the minimum of the free energy surface while a white noise term randomly drives these fields. The result is a Langevin equation for the dynamics of the fields
\begin{equation}
\label{Langevin}
\partial_t \phi = -\left( a \phi + 2b |\phi|^2 \phi - \nabla^2 \phi
\right) + \xi,
\end{equation}
where the noise correlations are those of white noise
\begin{equation}
\label{noise}
\langle\xi^*(x,t) \xi(x',t')\rangle = \frac{2}{\beta} \, \delta(x-x')
\delta(t-t').
\end{equation}
Note that this is a grand-canonical method, so the inverse temperature $\beta$ appears explicitly.

\subsection{The non-interacting case}

Let us first look at the non-interacting case ($b=0$) in which Eq.~(\ref{Langevin}) takes on the form of an Ohnstein-Uhlenbeck process and is analytically solvable \cite{Gardiner97}.

For long times, $t\rightarrow \infty$, the distribution function of field amplitudes $P(\textrm{Re}(\phi) = u)$ approaches
\begin{equation}
P(\textrm{Re}(\phi) = u) = \sqrt{\frac{\beta\sqrt{2a}}{\pi}} e^{-\beta\sqrt{2a}u^2},
\label{distri_analytic}
\end{equation}
and the spatial correlation function becomes
\begin{equation}
\langle\phi(\bullet)^* \phi(\bullet+r)\rangle =
\frac{1}{\beta\sqrt{2a}} e^{-\sqrt{2a}(|r|/\xi_0)},
\label{corr_analytic}
\end{equation}
which in momentum space reads
\begin{equation}
\langle |\phi_k|^2 \rangle = \frac{1}{\beta((\xi_0k)^2+2a)},
\label{corr_fourier_analytic}
\end{equation}
expressing the equipartition of energy between the classical modes with the free dispersion relation $E(k) = k^2$.

Using a second-order Runge-Kutta method, we propagate the initial state $\phi(x,0) = 0$ with the Langevin equation (\ref{Langevin}), where the stochastic noise (\ref{noise}) is interpreted in the It\^{o} sense \cite{Gardiner97}. After the observables, i.e.~the distribution and correlation functions, have settled to a steady state, we calculate their values by averaging over a set of trajectories and compare them to the analytic expressions in Eqs.~(\ref{distri_analytic}), (\ref{corr_analytic}) and (\ref{corr_fourier_analytic}). We used a uniform spatial grid with step size $\Delta x = 0.2$, an interval of length $L= 20$, a time step $\Delta t=0.05 \Delta x$ and the parameters $\beta = 2$, $a=1/2$, $b=0$, similar to the work in \cite{Lythe01}. We averaged over 1000 trajectories at a final time of $t_f = 20$, when the system has reached a steady state.

\begin{figure}
\centering
\textbf{(a)}\subfigure{\includegraphics[width=0.8\columnwidth]{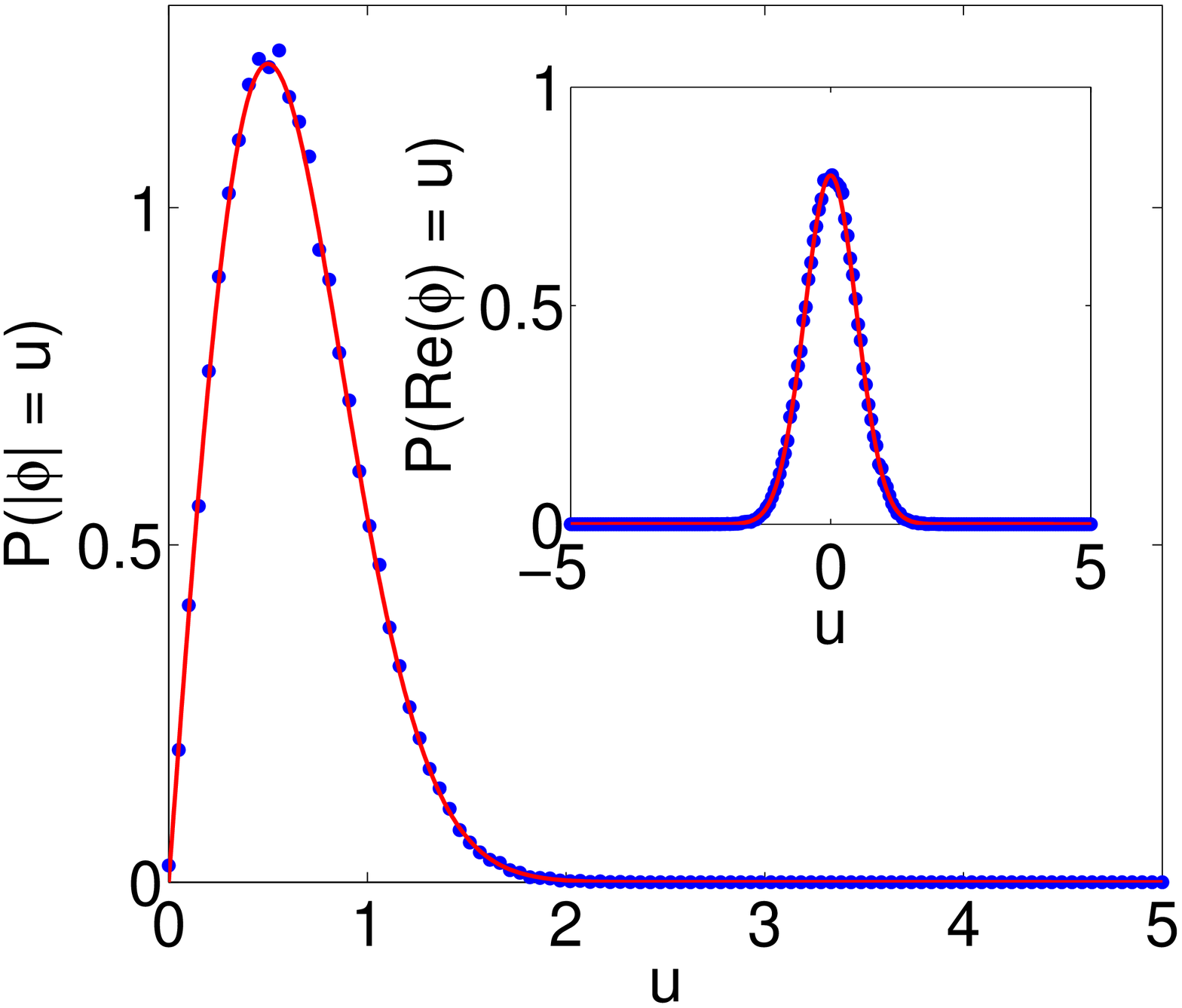}}
\textbf{(b)}\subfigure{\includegraphics[width=0.8\columnwidth]{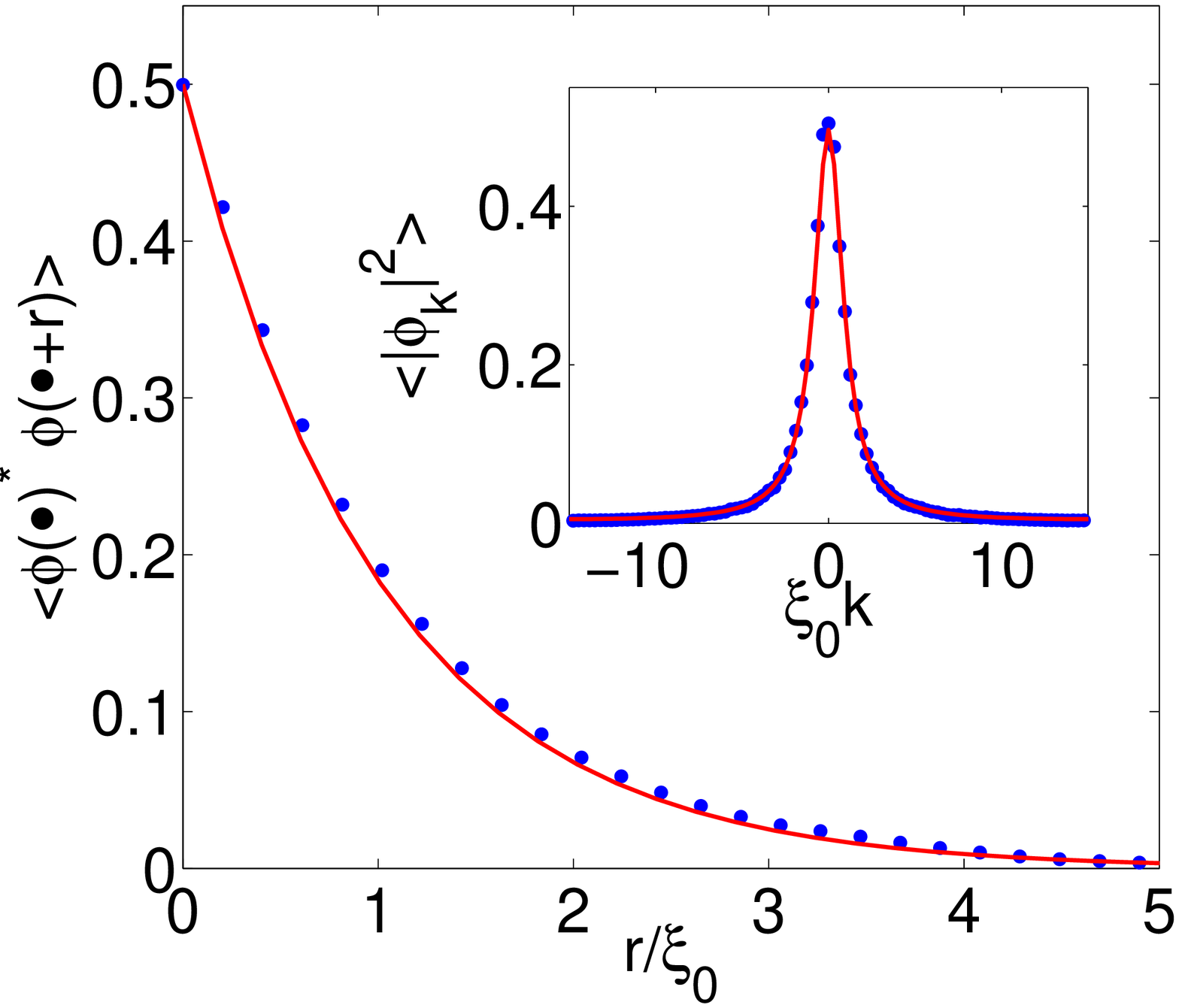}}
\caption{(Color online) Langevin dynamics in the non-interacting case ($\beta = 2$, $a=1/2$, $b=0$). \textbf{(a)} Radial distribution function $P(|\phi| = u)$ and (inset in \textbf{(a)}) Distribution function $P(\textrm{Re}(\phi) = u)$, \textbf{(b)} Correlation function in real space $\langle\phi(\bullet)^* \phi(\bullet+r)\rangle$ and (inset in \textbf{(b)}) in momentum space $\langle |\phi_k|^2 \rangle$. Numerical results (points) from the steady state of the Langevin equation (\ref{Langevin}) and analytic results from (\ref{distri_analytic}), (\ref{corr_analytic}) and (\ref{corr_fourier_analytic}) (solid line).}
\label{fig1}
\end{figure}

In Fig.~\ref{fig1} we show the results of our numerical simulations. The distribution function for the real part of the field $P(\textrm{Re}(\phi) = u)$ and the distribution function for the modulus of the complex field amplitude $P(|\phi| = u)$, as well as the correlation functions in real space $\langle\phi(\bullet)^* \phi(\bullet+r)\rangle$ and its Fourier transform $\langle |\phi_k|^2 \rangle$ are presented aside with the analytical solutions (\ref{distri_analytic}), (\ref{corr_analytic}) and (\ref{corr_fourier_analytic}).

The equilibrium properties of the numerical simulation of Eq.~(\ref{Langevin}) agree extremely well with the analytic solutions. We also used this baseline test to check the transfer integral approach, which overlaps exactly with the analytical solutions.

\subsection{The interacting case}

\begin{figure}
\centering
\textbf{(a)}\subfigure{\includegraphics[width=0.8\columnwidth]{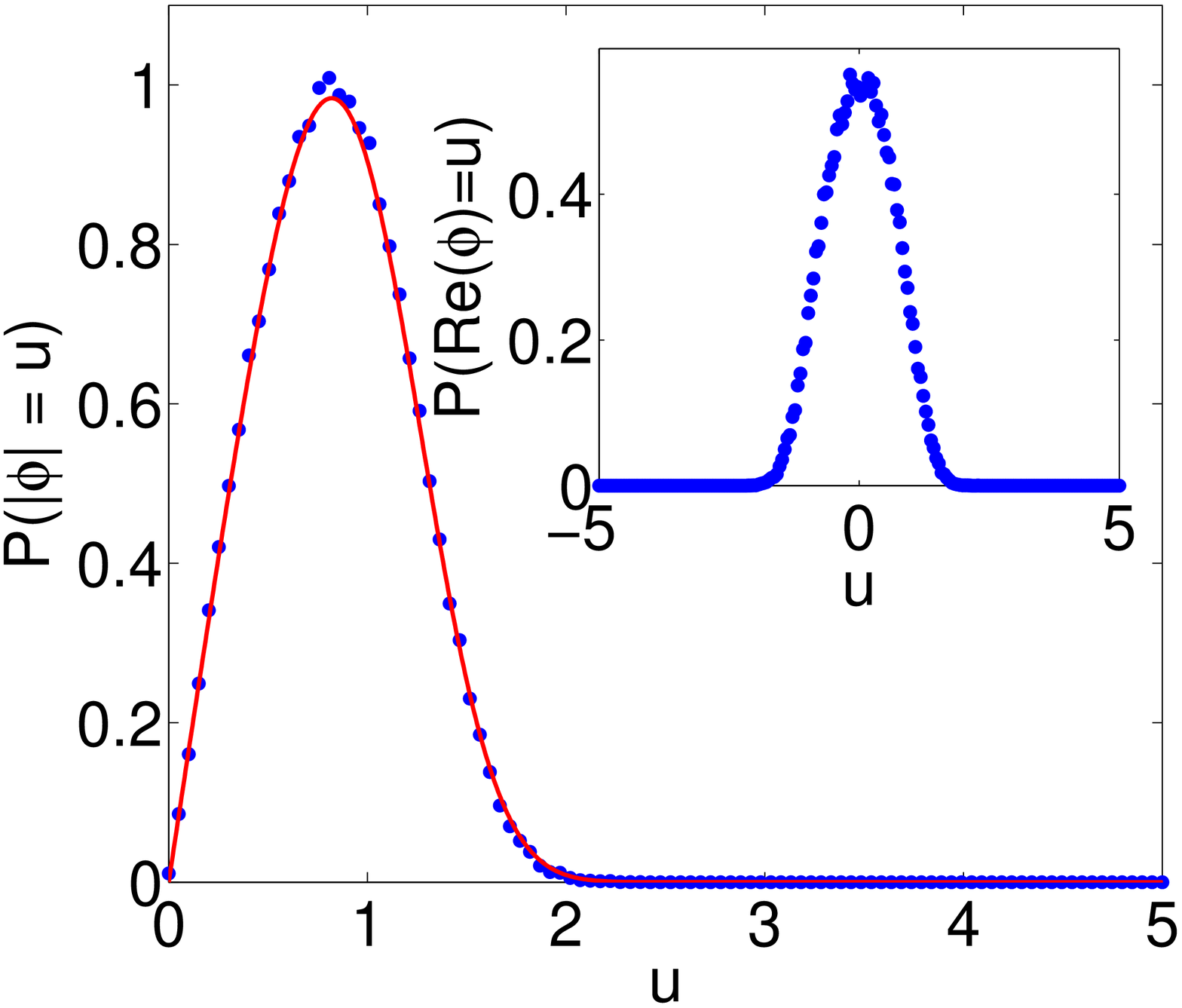}}
\textbf{(b)}\subfigure{\includegraphics[width=0.8\columnwidth]{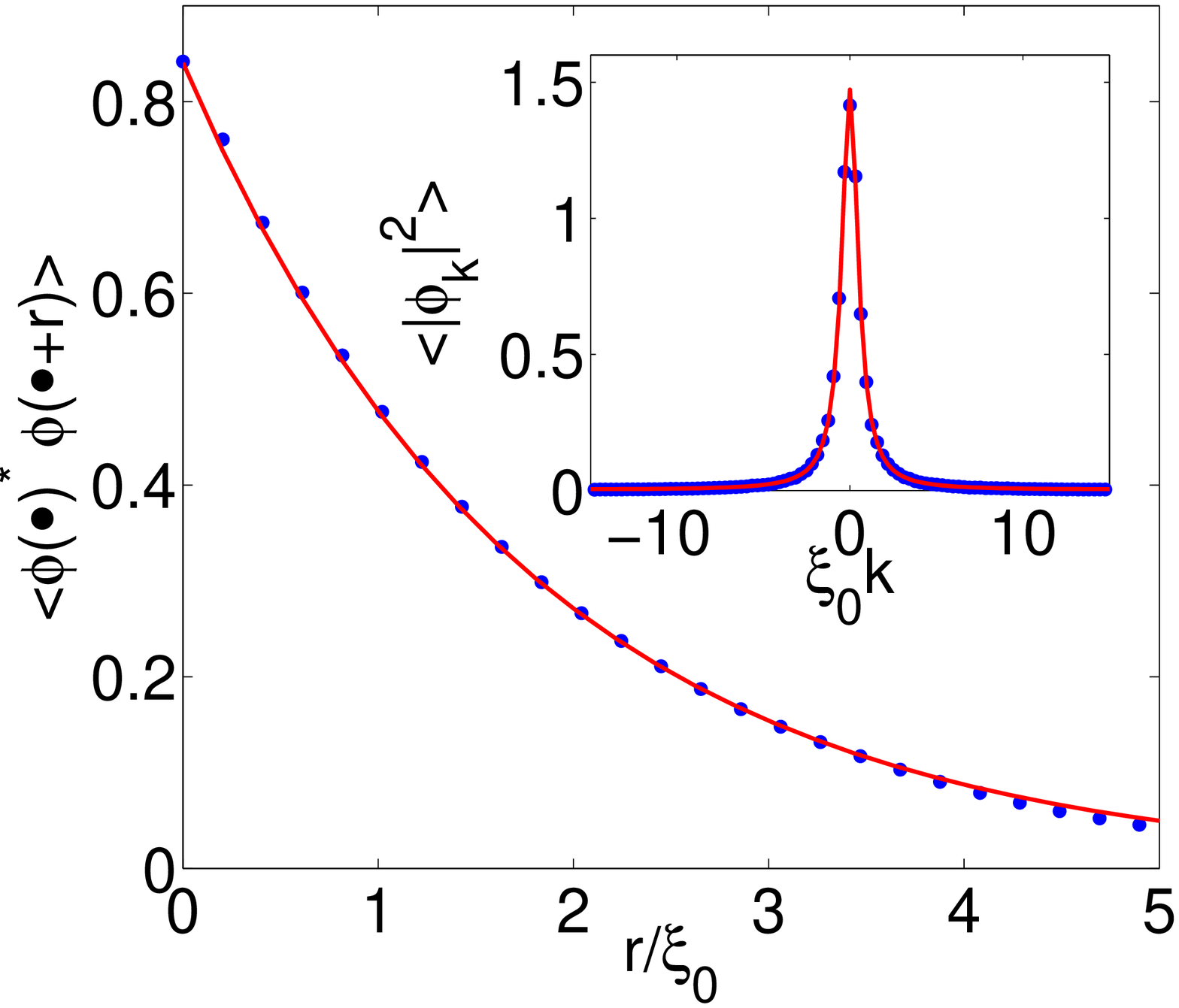}}
\caption{(Color online) Langevin dynamics in the interacting case ($\beta = 2$, $a=-1/2$, $b=1/4$). \textbf{(a)} Radial distribution function $P(|\phi| = u)$ and (inset in \textbf{(a)}) Distribution function $P(\textrm{Re}(\phi) = u)$, \textbf{(b)} Correlation function in real space $\langle\phi(\bullet)^* \phi(\bullet+r)\rangle$ and (inset in \textbf{(b)}) in momentum space $\langle |\phi_k|^2 \rangle$. Numerical results (points) from the steady state of the Langevin equation (\ref{Langevin}) and exact solution from the transfer integral method (\ref{transferintegral}) (solid line).}
\label{fig2}
\end{figure}

With the transfer integral method we can extend our studies to the non-linear case ($b \not= 0$), where results cannot be obtained analytically. In Fig.~\ref{fig2} we plot the distribution and correlation functions for $\beta = 2$, $a=-1/2$ and $b=1/4$. All other parameters were the same as before. We see that the stochastic simulation correctly reproduces the exact thermodynamic results for this interacting system in the limit of long times where the system has settled to a steady state.

\begin{figure}
\centering
\textbf{(a)}\subfigure{\includegraphics[width=0.8\columnwidth]{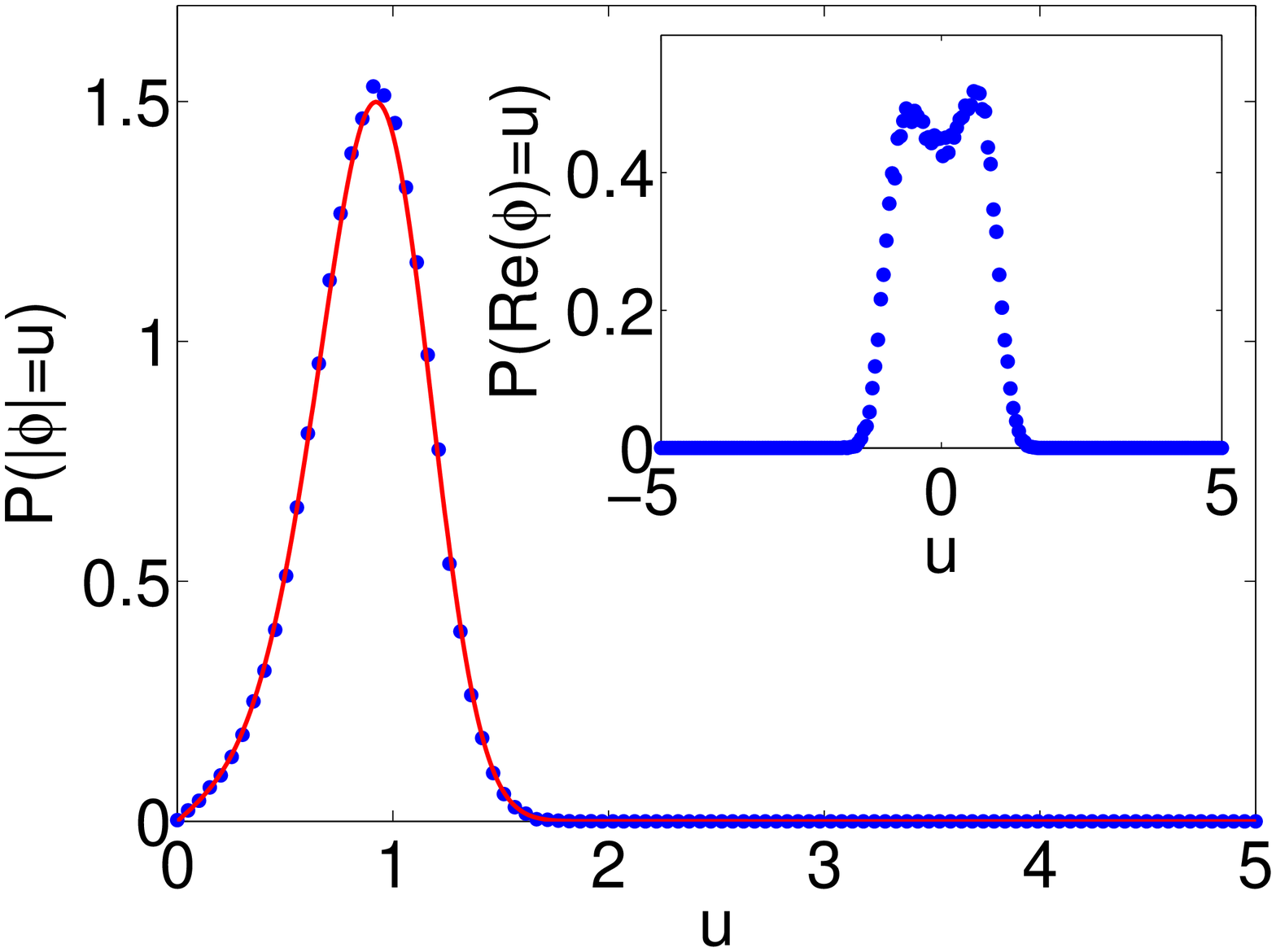}}
\textbf{(b)}\subfigure{\includegraphics[width=0.8\columnwidth]{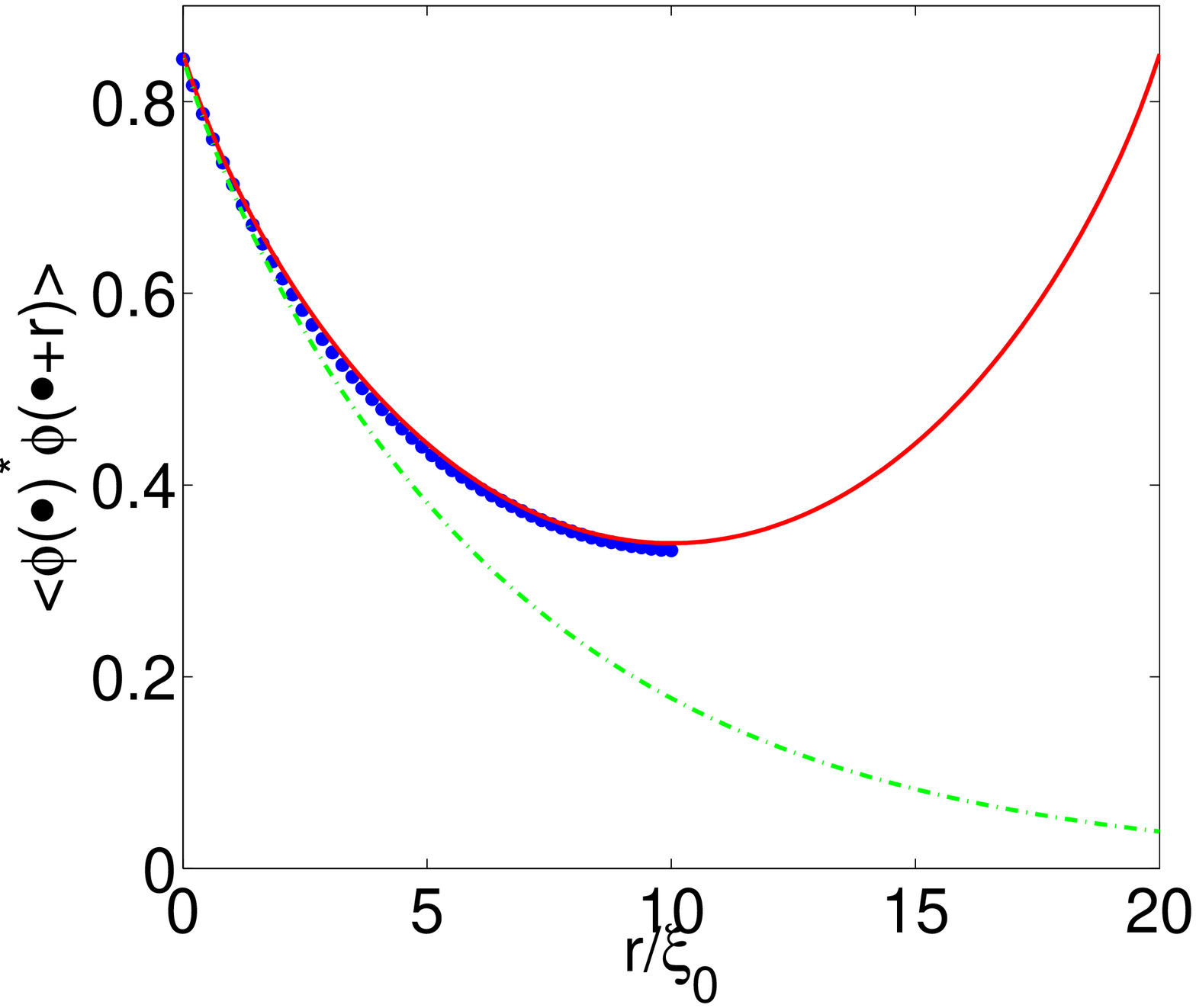}}
\caption{(Color online) Langevin dynamics in the interacting case ($\beta = 6$, $a=-1/2$, $b=1/4$). \textbf{(a)} Radial distribution function $P(|\phi| = u)$ and (inset in \textbf{(a)}) Distribution function $P(\textrm{Re}(\phi) = u)$. Numerical results (points) from the steady state of the Langevin equation (\ref{Langevin}) and exact solution from the transfer integral method (\ref{transferintegral}) (solid line). \textbf{(b)} Correlation function in real space $\langle\phi(\bullet)^* \phi(\bullet+r)\rangle$ aside with numerical results (points). The transfer integral solution is presented including finite-size corrections (\ref{corr_function_finite}) (solid line) and in the thermodynamic limit (\ref{corr_function}) (dashed line).}
\label{fig3}
\end{figure}

Fixing all parameters, we explore the equilibration process as we decrease the temperature, i.e.~increase $\beta$. In Fig.~\ref{fig3} distribution and correlation functions are shown for $\beta = 6$. We find that the time to reach the steady state increases by over an order of magnitude, from $t_f = 20$ to $t_f=250$. In the real part of the distribution function we can identify the minima of the Mexican hat potential $V(x) = a x^2 + bx^4$ to which the dynamics is limited for small temperatures. The distribution function $P(|\phi|=u)$ still agrees well with the exact solution (\ref{distribution}). While the simulation equilibrates to the correct density $\langle |\phi(\bullet)|^2 \rangle$, its spatial correlation function $\langle \phi(\bullet)^* \phi(\bullet+r) \rangle$ deviates for large spatial scales from the transfer integral as given in Eq.~(\ref{corr_function}).

\begin{figure}
\centering
\includegraphics[width=0.8\columnwidth]{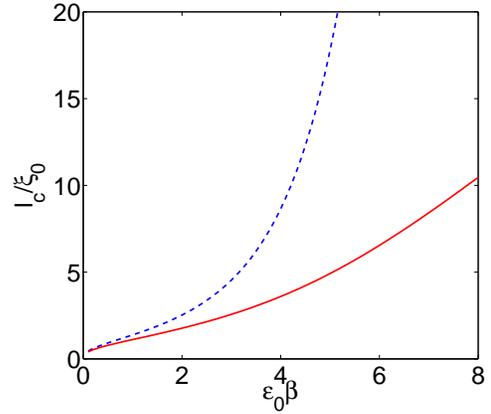}
\caption{(Color online) Correlation length $l_c$ as a function of the inverse temperature $\beta$ for a complex field (solid line). Additionally, we plot the correlation length $l_c$ for a real field (dashed line). For real fields, the correlation length increases much more dramatically with decreasing temperature and, therefore, equilibration at temperatures $\beta>6$ requires very long interaction times.}
\label{fig4}
\end{figure}

To explain this behavior we recall that the long-range properties of the correlation function are determined by the
two smallest eigenvalues of the transfer integral:  $E_0 < E_1$. We therefore define a correlation length $l_c$ by
\begin{equation}
\frac{l_c}{\xi_0} = \frac{1}{\beta(E_1 - E_0)}.
\end{equation}
In Fig.~\ref{fig4} the correlation length $l_c$ is calculated from the transfer integral of Eq.~(\ref{transferintegral}) as a function of the inverse temperature $\beta$. The correlation length $l_c$ becomes comparable to the system size $3l_c \approx L/2$ for $\beta > 4$. At this temperature coherence is established across the whole system which is not accounted for by our expression for the correlation function as given by Eq.~(\ref{corr_function}). Instead, we must properly account for the finite size of the ring in our derivation of the correlation function (which we show in the appendix). Using Eq.~(\ref{corr_function_finite}), we plot in Fig.~\ref{fig3} the exact result for the correlation function of this finite system. The simulation data agrees very well with this version of the correlation function. Moreover, we note that the long equilibration times observed in the simulations at lower temperatures are explained by the fact that it takes many collisions to establish coherence between the particles, the typical 'slowing down' phenomenon close to a phase transition.

\section{Molecular dynamics}

One of the disadvantages of the stochastic method we have just presented is that it is difficult to generalize to an inhomogeneous situation. Here we present an alternative CF method which (1) may easily accommodate an inhomogeneous potential, (2) is a microcanonical method so may be more directly compared to the classical field method, and for which (3) an effective temperature of the interacting system may be unambiguously determined because the momentum field is decoupled from the density field.

Starting from the partition function of Eq.~(\ref{classpf}), one calculates observables by the relation
\begin{equation}
 \langle \hat O \rangle = \frac{\int D(\phi,\phi^*) O(\phi,\phi^*)
 e^{-S}}{\int D(\phi,\phi^*) e^{-S}}.
\end{equation}
We introduce the canonical momentum field $\pi = \partial_t \phi$, which is independent of the density field $\phi$, by
multiplying the partition function of Eq.~(\ref{classpf}) by unity
\begin{equation}
 \int D(\pi,\pi^*) e^{-|\pi|^2} \propto 1
\end{equation}
and arrive at the partition function
\begin{equation}
Z \propto \int D(\pi,\pi^*) \int D(\phi,\phi^*) \exp{ \left[ -\beta
\int dx \, \tilde H[\pi,\phi] \right]},
\end{equation}
where $\tilde H[\pi,\phi]$ may be treated as a Hamiltonian with the action $S[\phi]$ playing the role of a potential
\begin{equation}
\tilde H[\pi,\phi] = |\pi|^2 + S[\phi].
\end{equation}
We may readily derive Hamilton's equations of motion
\begin{eqnarray}
\dot \pi &=& - \left( a \phi + 2b |\phi|^2 \phi - \nabla^2 \phi
\right), \\
\dot \phi &=& \pi,
\end{eqnarray}
and combine them to obtain
\begin{equation}
\partial_t^2 \phi = \nabla^2 \phi - a \phi - 2b |\phi|^2 \phi.
\label{molecular}
\end{equation}

Using a standard leap-frog integration routine we solve Eq.~(\ref{molecular}) numerically for the non-equilibrium initial
condition
\begin{equation}
 \phi(x,0) = 0, \hspace{1cm} \pi(x,0) = \sum_{k=1}^{N_\mathrm{ex}} A
 e^{2\pi i (kx/L - \xi_k)},
 \label{md_init}
\end{equation}
where $k$ is an integer and $\xi_k$ are random numbers drawn from a uniform distribution of $0 \le \xi_k < 2\pi$. The initial energy is then simply $E = N_\mathrm{ex} L A^2$ \cite{Aarts00}. For our simulation, we choose, similar to the last section, $N_{\text{ex}} = 4$, $E/N = 2$, $a = -1/2$, and $b = 1/4$ as well as $L = 20$, $\Delta x = 0.2$, and $\Delta t = 0.05 \Delta x$.

\begin{figure}
\centering
\includegraphics[width=0.8\columnwidth]{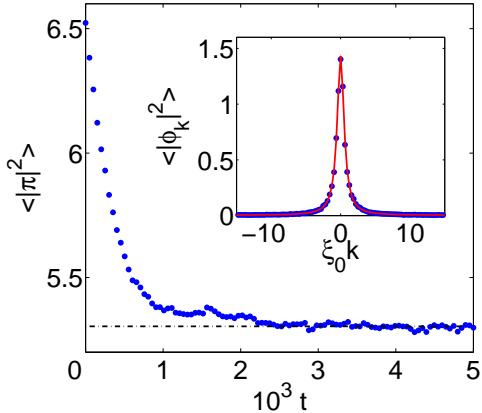}
\caption{(Color online) Molecular dynamics. Kinetic energy per particle $\langle |\pi|^2 \rangle$ versus time $t$ and (inset) correlation function in Fourier space in the steady state $\langle|\phi_k|^2\rangle$. Simulation results (points) and exact solution (solid line).}
\label{fig5}
\end{figure}

In Fig.~\ref{fig5} we show the average kinetic energy per degree of freedom (number of grid points in the simulation) as a function of time. The system comes to a steady state after mixing for $2500$ time units. In the steady state, we compute the correlation function $\langle|\phi_k|^2\rangle$ by replacing ensemble averages with time averages over $500$ time units and, additionally, over 10 trajectories, to further reduce fluctuations.  We may now determine an effective
temperature $T_{\textrm{eff}}$ of the interacting system by computing the time-averages of the kinetic energy $\langle |\pi|^2 \rangle$
\begin{equation}
T_{\textrm{eff}} = \langle |\pi|^2\rangle \, \Delta x/2 = 0.47.
\end{equation}

In the inset of Fig.~\ref{fig5} we plot the correlation function from our simulations aside with the exact solution obtained from the transfer integral approach for $\beta = 1/T_{\textrm{eff}}$. The agreement between the properties of the steady state and the exact solution, as well as with the simulations of the previous section, is very good.

We would like to stress that since the $\pi$ field is decoupled from the density field $\phi$, a temperature may be reliably determined from these microcanonical simulations.  This, unfortunately, is not the case in the following microcanonical method.

\section{``Classical Field Method''}

Let us now approach the interacting Bose gas from a different perspective. To begin, we will start from the real-time quantum action for an interacting Bose gas in the microcanonical ensemble
\begin{equation}
S[\phi,\phi^*] = -\phi^* i\partial_{t} \phi + b |\phi|^4 + |\nabla
\phi|^2.
\end{equation}
Minimizing the action, which is equivalent to making a saddle-point approximation to the full quantum mechanical kernel, we arrive at
\begin{equation}\label{GPE}
0 = \frac{\delta S}{\delta \phi^*} = -i\partial_t \phi + 2b|\phi|^2
\phi - \nabla^2 \phi,
\end{equation}
i.e.~the usual non-linear Schr\"odinger equation (NLSE). At this level of approximation the dynamics of the fields are, roughly speaking, restricted to classical paths, and one might think that a microcanonical ensemble of particles, governed by such an equation of motion, displays equilibrium properties determined by the classical partition function of Eq.~(\ref{classpf}).

\begin{figure}
\centering
\includegraphics[width=0.8\columnwidth]{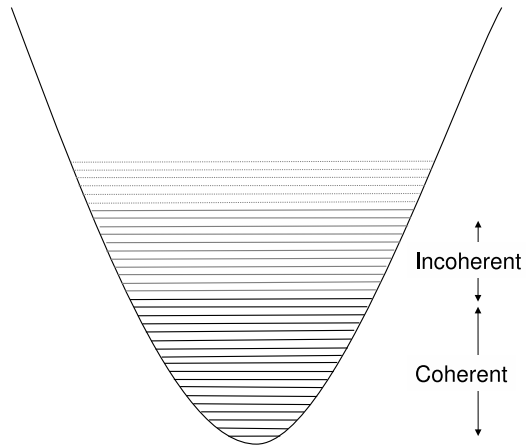}
\caption{Schematic diagram showing the classical and incoherent regions of the single particle spectrum for a harmonically trapped Bose gas.}
\label{fig6}
\end{figure}

Note, the NLSE equation is usually thought to account only for the lowest energy mode of a Bose gas, but this restriction is nowhere imposed in the derivation above. The validity of the saddle-point approximation, however, will depend upon the extent to which each mode can be treated classically. In Fig.~\ref{fig6} we depict the energy levels of a harmonically trapped Bose gas. In a finite region, from the ground state up to some cutoff energy, all the modes are multiply occupied.  In this region, quantum fluctuations are negligible and we may treat these particles as a set of coherent or classical fields. For more details as to the subtleties of this method, such as how to incorporate the incoherent portion of the spectrum, we refer the reader to the literature \cite{Davis01,Goral01,Sinatra01,Davis05}.

We start our simulation from the non-equilibrium state
\begin{equation}
\phi(x) = \sum_{n=-N_{\textrm{ex}}}^{N_\textrm{ex}} \frac{e^{2\pi i n
x}}{\sqrt{L}} e^{i\xi_n},
\end{equation}
where $2N_\textrm{ex}+1$ is the number of excited modes, $L$ the system length and $\xi_n$ are random numbers drawn from a uniform distribution in the interval $[0,2\pi]$. We propagate this initial condition with the NLSE (\ref{GPE}) using a standard semi-spectral fourth-order Runge-Kutta routine with $L = 20$, $\Delta x = 0.2$ and $\Delta t = 10^{-4}$ and choose the non-linear coefficient to be $C = 2b = 1000$. After the NLSE (\ref{GPE}) reaches a steady state \cite{Davis01}, we take time averages over a period of 1000 time units to calculate the momentum distribution $\langle |\phi_k|^2 \rangle$. To compare with the exact transfer integral solution, we first adjust the chemical potential $a$ to match the particle number. Unfortunately, we do not know of a simple way to directly extract the temperature from these simulations when the interactions are large, as in the case we are currently concerned with. However, we may avoid this issue by fitting the curves to the exact solution, treating the the temperature $\beta$ as a free parameter.

\begin{figure}
\centering
\includegraphics[width=0.8\columnwidth]{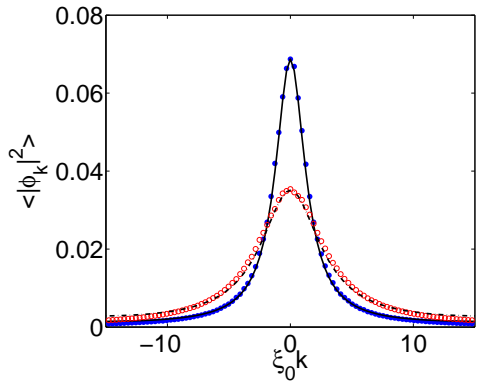}
\caption{(Color online) Classical Field Method. Steady-state momentum distribution at $C = 2b = 1000$ (points) and the new steady-state after the change in the interaction constant to $\tilde C = 2 \tilde b =  5000$ (circles). Both are fitted to the exact solution given by the transfer integral (\ref{transferintegral}) (solid and dashed line).}
\label{fig7}
\end{figure}

In Fig.~\ref{fig7} we show the momentum distribution of the steady state, to which the classical field method equilibrates at strong non-linearities, and compare it to an exact solution for $\beta = 5.29$, $a = -63.1$ and $b = 1000/2$. The agreement is excellent which strongly supports the conclusion that the classical field method is able to correctly sample the classical partition function.

If, after equilibrium has been reached, we suddenly change the interaction constant to a different value, $\tilde C = 2 \tilde b = 5000$, we find that the momentum distribution changes to the new equilibrium value which can be fitted with the exact solution $\beta = 2.55$, $a = -313.3$ and $\tilde b = 5000/2$. We note that the fit does not reproduce the tails of the distribution exactly, however, the overall fit is still excellent.

We believe that these observations are a strong motivation for using the ``classical field method'', since it is based upon the NLSE, to describe dynamics of cold atomic gases at finite temperatures. This subject, however, is beyond the scope of the current paper.

\section{Conclusion and Outlook}

We have compared three different CF approaches to dynamically sample the classical partition function to an exact transfer integral solution. First, a stochastic dynamical method, based upon the evolution of a Langevin equation, was shown to contain the correct thermal fluctuations needed to recover the exact distribution and correlation functions of the full classical theory. Second, we presented the microcanonical method of molecular dynamics, which was again able to reproduce the exact one-dimensional solution. Both of these dynamical schemes have either a temperature built-in or one which may easily be determined -- although the latter is more easily adapted to an inhomogeneous system. Third and final, we have evaluated an alternative CF method, the ``classical field method'', and shown that its steady state is given by the exact result for the partition function of a Bose gas in the high-temperature limit.

Furthermore, when we changed the interaction strength of our ``classical field method'' simulations, the system was seen to dynamically evolve to another equilibrium point even in this low-dimensional system. This observation provides a strong motivation that the classical field method may be used to study finite-temperature dynamics. It would be especially interesting to apply this formalism to experiments in 1D ring geometries with Josephson junctions which can be realized e.g.~by perturbing the homogeneous ring by a potential barrier. In this case, the molecular dynamics approach should be valuable as it can easily be extended to this inhomogeneous situation and provides an effective temperature for the interacting system, while the classical field method offers a powerful means to study the dynamical relaxation process. Another line of research might be the study of heat conduction in transport experiments on strongly-interacting systems.

\appendix*
\section{Derivation of the transfer integral solution}

In this appendix we review the evaluation of the classical partition function, Eq.~(\ref{classpf}), by means of the transfer integral \cite{Scalapino72}.  Assuming a free energy of the form of Eq.~(\ref{free_energy}), we next discretize the one-dimensional domain $[0,L]$ with $N$ equally spaced grid points and apply periodic boundary conditions $\phi_{N+1} = \phi_1$, turning the domain into a ring of length $L$. The partition function then takes the form
\begin{equation}
Z = \prod_{i=1}^N \int d\phi_i \, e^{ - \beta (\Delta x/\xi_0)
f(\phi_{i+1},\phi_i)},
\end{equation}
where
\begin{equation}
f(\phi_{i+1},\phi_i) = a |\phi_{i+1}|^2 + b |\phi_{i+1}|^4 +
\left|\frac{\phi_{i+1}-\phi_i}{\Delta x}\right|^2.
\end{equation}
By introducing the following generalized notation
\begin{equation}
\int d\phi_{i} \, e^{ - \beta f(\phi_{j},\phi_i) } |\phi_i \rangle
\rightarrow \sum_i T_{j,i} |\,\rangle_i \rightarrow \hat T |\,\rangle,
\end{equation}
the partition function can be rewritten as
\begin{equation}
Z = \mathrm{tr} \left[ \hat T^N \right].
\end{equation}
If we can find the eigenfunctions $|\varphi_n\rangle$ and eigenvalues $E_n$ of the transfer operator $\hat T$, the partition function can be exactly evaluated through the relation
\begin{equation}
Z = \sum_{n=0}^\infty e^{- \beta (L/\xi_0) E_n}.
\end{equation}
The eigenvalue problem $\hat T |\varphi_n\rangle = \lambda_n |\varphi_n\rangle $ reads explicitly
\begin{equation}
\int d\phi_{i} \, e^{ - \beta f(\phi_{i+1},\phi_i)} \varphi_n(\phi_i)
= e^{-\beta E_n} \varphi_n(\phi_{i+1}),
\end{equation}
and can be restated by expanding $\varphi_n(\phi_i)$ in a Taylor series about $\varphi_n(\phi_{i+1})$ as
\begin{equation}
 e^{ - \beta H } \varphi_n = e^{ - \beta E_n } \varphi_n,
\end{equation}
where $H$ is the Hamiltonian of a single particle in an anharmonic oscillator potential
\begin{equation}
H = -\frac{1}{4\beta^2} \nabla_u^2 + a |u|^2 + b |u|^4.
\end{equation}
In this way we have reduced the problem of calculating the classical partition function (\ref{classpf}) to finding the eigenvalues and eigenfunctions of a simple, one-particle, time-independent Schr\"odinger equation $H \varphi_n = E_n \varphi_n$.

The spatial correlation function $\langle \phi(\bullet)^* \phi(\bullet+r)\rangle$ can be determined as follows
\begin{eqnarray}
\langle \phi(\bullet)^* \phi(\bullet+r)\rangle \nonumber &
=& Z^{-1} \int D(\phi,\phi^*) \phi(0)^* \phi(r) e^{-\beta F} \nonumber \\
&=& Z^{-1} \, \mathrm{tr} \left[ \phi(0) \hat T^r \phi(r) \hat T^{N-r} \right].
\end{eqnarray}
Using the spectral representation of $\hat T = |\varphi_n\rangle \lambda_n \langle \varphi_n|$, we obtain
\begin{eqnarray}
\label{corr_function_finite}
&&\langle \phi(\bullet)^* \phi(\bullet+r)\rangle \nonumber\\
&=&\frac{ \sum_{ij} \langle \varphi_j | \phi(0) | \varphi_i \rangle
\left( e^{-\beta E_i} \right)^r \left( e^{-\beta E_j} \right)^{N-r}
\langle \varphi_i | \phi(r) | \varphi_j \rangle}
{ \sum_n \left( e^{-\beta E_n} \right)^{N}} \nonumber\\
&=& \frac{\sum_{ij} |\langle \varphi_i | \phi | \varphi_j \rangle|^2
\left( e^{-\beta E_i} \right)^r \left( e^{-\beta E_j} \right) ^{N-r}}
{ \sum_n \left( e^{-\beta E_n} \right)^{N}},
\end{eqnarray}
where we have used the spatial homogeneity of the system to arrive at the last line.

In the thermodynamic limit, $N \rightarrow \infty$, and when the system size is large compared to the coherence length $l_c\ll L$, we can simplify this formula to
\begin{equation}
\langle \phi(\bullet)^* \phi(\bullet+r)\rangle = \sum_{i=1}^\infty
|\langle \varphi_i | \phi | \varphi_0 \rangle|^2 e^{-\beta (E_i-E_0)
(r/\xi_0)},
\end{equation}
where the matrix element has to be evaluated through
\begin{equation}
\langle \varphi_i | \phi | \varphi_0 \rangle = \int_0^\infty du \, \varphi_i^*(u) \, u \, \varphi_0(u).
\end{equation}

\acknowledgments A. N.~acknowledges a scholarship from the Rhodes
Trust, J. N. M.~a USA research fellowship from the Royal Society, and
K. B.~support from the Royal Society and the Wolfson Foundation.

\end{document}